# *THE PICTURE OF DORIAN GRAY WITH ROLES REVERSED: NOTES FOR A CHRONOLOGY OF THE TELESCPE-MAKING ACTIVITIES OF THE NEAPOLITAN OPTICIAN FRANCESCO FONTANA (C. 1585–1656)*

Paolo Del Santo
Museo Galileo: Institute and Museum of History of Science – Florence (Italy)
E-mail: p.delsanto@gmail.com

**Abstract:** In the last few years it has been alleged that the sitter in Jusepe de Ribera's Allegory of Sight painting is the Neapolitan optician Francesco Fontana (ca. 1585–1656), who is well-known for contributing in a decisive manner to the diffusion of the Keplerian telescope. The present paper demonstrates the impossibility of this identification, based mainly on the erroneous assumption that Fontana was already renowned as an optician by the mid-1610s. Instead, we suggest a more reasonable chronology for his activities, which postpones the spreading of his fame out of Naples until the end of the following decade.

**Sommario:** Da alcuni anni è stata avanzata l'ipotesi che l'uomo raffigurato nell'*Allegoria della vista* di Jusepe de Ribera possa essere identificato con l'ottico napoletano Francesco Fontana (c. 1585–1656), noto per aver contribuito in maniera determinante all'affermazione del telescopio kepleriano. Nel presente articolo si dimostra l'impossibilità di tale identificazione, basata sull'assunto erroneo che Fontana fosse un ottico affermato già nella metà degli anni '10 del XVII secolo, e si suggerisce una più convincente cronologia della sua attività, secondo la quale la sua fama si diffuse al di fuori della città di Napoli solo alla fine del decennio successivo.

**Keywords**: Francesco Fontana, history of the telescope, early telescopes, Jusepe de Ribera, Fabio Colonna

## 1 THE TWO PORTRAITS

In a paper appeared in 2017, Paolo Molaro (2017b: 284-86), an Italian astrophysicist at the Astronomical Observatory of Trieste, analyses the life and the work of the Neapolitan optician Francesco Fontana (c. 1585–1656), known for contributing in a decisive manner to the emergence and diffusion of the Keplerian (or astronomical) telescope, i.e. with converging ocular. Among others, in the paper Molaro maintains the possible identification of the sitter in the painting *Allegory of Sight* (figure 1) by the famous Spanish Tenebrist painter Jusepe de Ribera (1591–1652), known in Italy as Lo Spagnoletto, with Fontana (Molaro, 2017b: 284-86).[1]

This speculation does not originate from one or more previously unknown sources, or from the reinterpretation of already known ones, but only from a visual impression (which would not deleterious in and of itself): "We note here that the sitter in *The Sight* by Ribera bears a close resemblance to the self-portrait made by Fontana for his book[2] [figure 2]" (Molaro, 2017b: 284). However, Molaro himself acknowledges that it is rather problematic to support, *sic et simpliciter*, this alleged resemblance:

> The shape of the head and the characteristics of the face and of the gaze are strikingly similar. One main difference between the two portraits lies in the hair. However, Fontana in 1646 presented himself as he looked in 1608 (i.e. almost 40 years younger), and the simplest way to look younger is by adding hair. Anyway, the possible Fontana in the painting by Ribera should be a few years older. Also, the ears are different, but it must be considered that Fontana's self-portrait cannot be compared to those of one of the most talented painters of his times. Thus, although it is generally believed that Ribera took his models from everyday life, it cannot be excluded that for the specific subject of the *Allegory of Sight* Ribera took inspiration from the figure of Fontana, who by this time was already a renowned telescope-maker. [...] A telescope decorated with gold is not something that can be associated with a man from the street since at that time it was very precious and was a symbol of power. We admittedly prefer the possibility that the man in Ribera's portrait could be the inventor of the *astronomical* telescope. (Molaro, 2017b: 286)

Actually, despite what Molaro may claim, the facial features of the two men are anything but similar, and the two faces look alike in the same way two random faces, seen head-on, look alike: two eyes, a nose and a mouth. The man portrayed by Ribera, for instance, has an evident aquiline nose, definitely different from that of Fontana's self-portrait, and the eye shape is quite different too. In any case, by admission of Molaro himself, the two portraits show indisputable, significant differences, at least in the ears and, above all, in the hair.

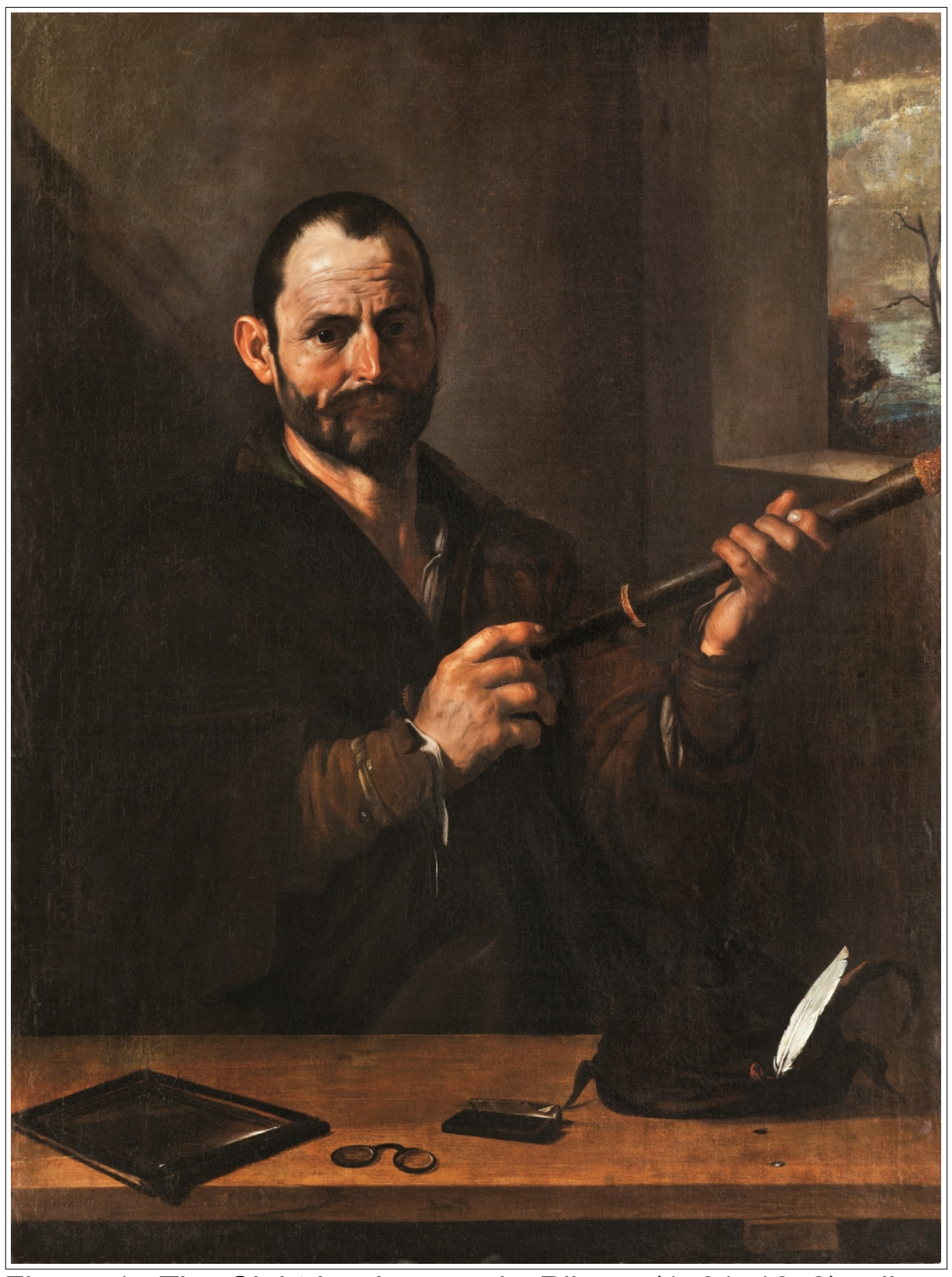

Figure 1: *The Sight* by Jusepe de Ribera (1591–1652), oil on canvas, 114 cm x 89 cm (courtesy: Franz Mayer Museum, Mexico City).

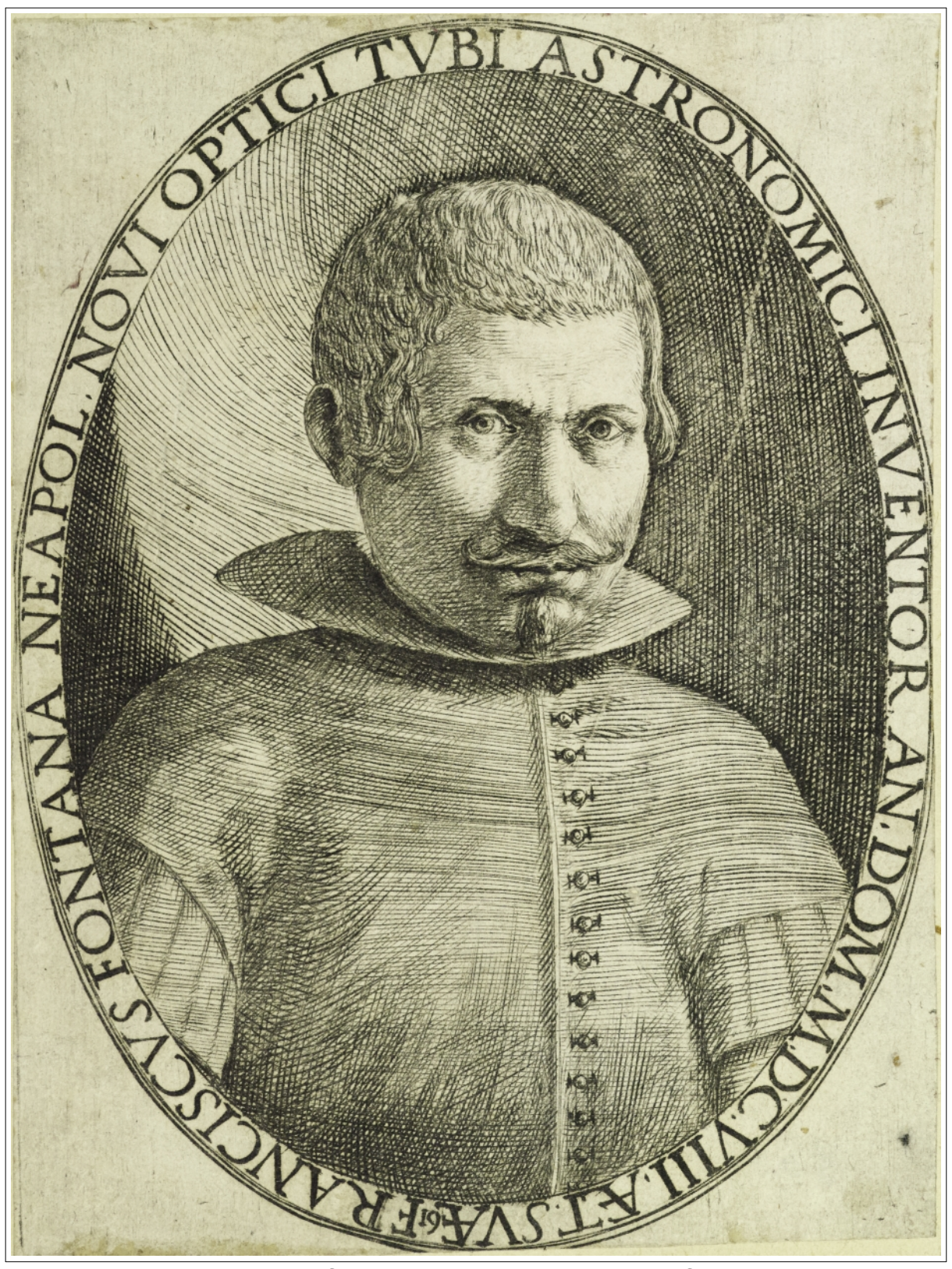

Figure 2: Portrait of Francesco Fontana from the *Novae cœlestium terrestriumque rerum observationes* (1646) (courtesy: Deutsche Museum von Meisterwerken der Naturwissenschaft und Technik, München).

As for the ears, we have seen that Molaro attributes the difference to Fontana's poor pictorial skills. However, a first question arises: if Fontana (or whoever made his portrait[3]) was not so good in drawing, then what is the reliability of his self-portrait? In other words, since Fontana's depiction might diverge significantly from his true face, does it make sense to compare the two representations?

The issue of the hair is definitely much more problematic: the receding hairline of Ribera's sitter differs markedly from the full head of hair of the self-portrait. This is rather odd, since the former was made thirty or more years before the latter, and not the reverse. How can we explain that? According to Molaro (2017: 286), such a difference between the two portraits would be due to Fontana's attempt to represent himself as he looked almost 40 years before, i.e. at the time of his alleged invention of the astronomical telescope, in 1608. In other words, Fontana would have thickened the hair of his self-portrait on purpose to represent himself about 40 years younger.

What does Molaro base his statement on? Actually, he offers no explanation for this, unless —but he does not expressly declare— this conjecture lies on a misinterpretation, or rather on a stretch of interpretation of the inscription around the portrait, which reads: "FRANCISCVS FONTANA NEAPOL[ITANVS] NOVI OPTICI TVBI ASTRONOMICI INVENTOR AN[NO] DOM[INI] MDCVIII ÆT[ATIS] SVÆ 61". The most obvious way to translate it is: "Francesco Fontana, from Naples, inventor, in the year 1608, of a new [kind of] astronomical telescope, at the [current] age of 61". However, in the past, some authors suggested the possibility of a different interpretation. In a short essay published at the beginning of the last century, Antonio Favaro, the eminent scholar, editor of the National Edition of Galileo's works, admits the possibility that the two digits indicating Fontana's age could be read upside-down, i.e. as "19" instead of "61" (Favaro, 1992: 424).[4] But in this case —since, for obvious reasons, Fontana could not be only nineteen years old in 1646, when he published his *Observationes*— “19” would necessarily refer to the age he was in 1608. In other words, although Favaro does not state it explicitly, if we read “19” instead of “61”, we must translate the inscription as follows: “Francesco Fontana, from Naples, inventor, in the year 1608, when he was 19 years old, of a new [kind of] astronomical telescope”. Maybe it was these same doubts around thirty years earlier that caused Pietro Riccardi, (1870: 467) in his famous *Biblioteca matematica*, to declare that he could not infer Fontana’s date of birth from the inscription surrounding his portrait.

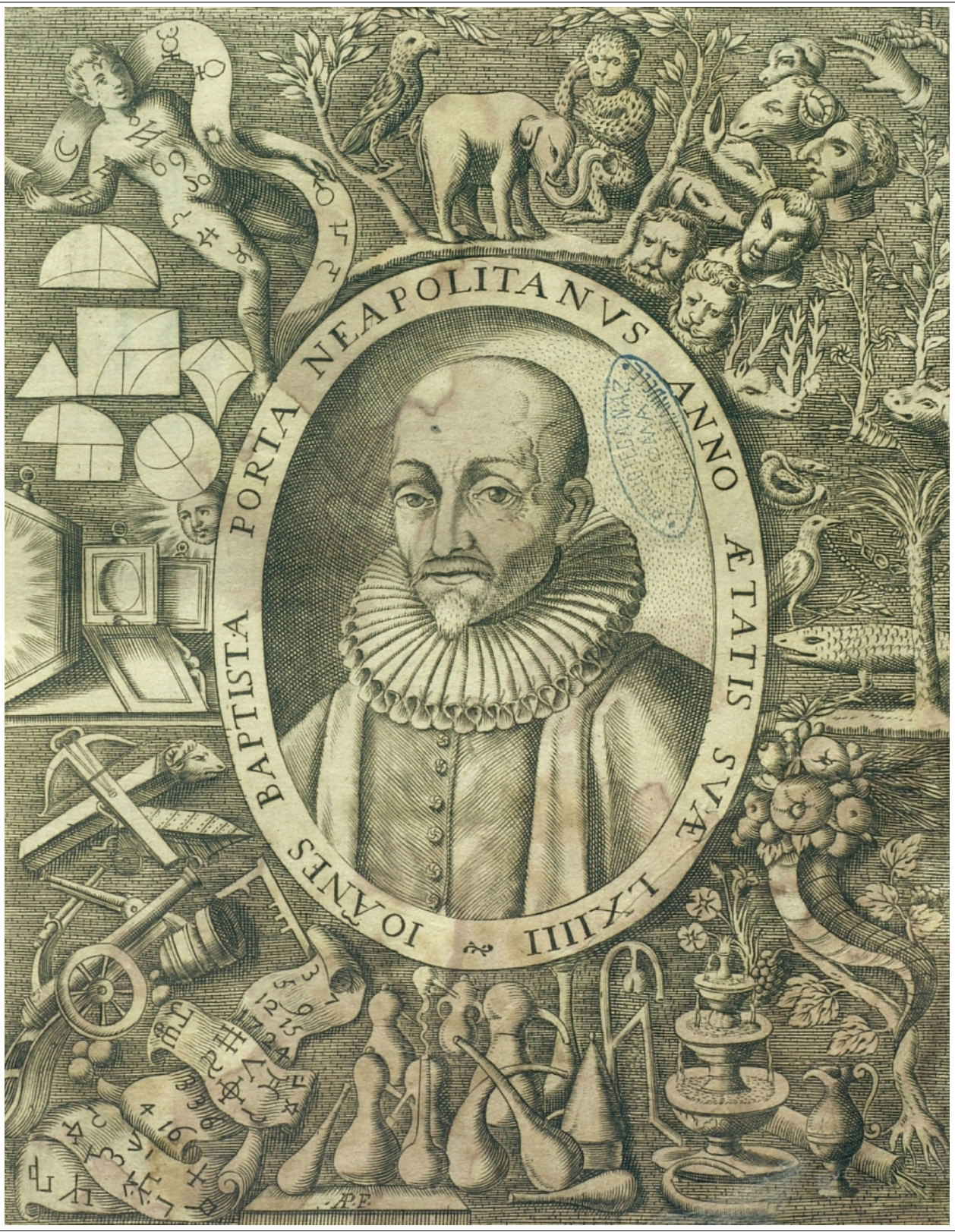

Figure 3: Portrait of Giovanni Battista Della Porta from the frontispiece of the *Magia naturale* (Della Porta, 1677) (courtesy: Biblioteca Nazionale Centrale di Roma))

In my opinion, incidentally, there is no special reason why the number indicating the age must be read upside down in relation to the immediately preceding and following words, as attested by other examples of similar portraits, in which the age of the person, written in the lower part of the oval frame, cannot be misunderstood. We see an example of this in the inscription surrounding the portrait of Giovanni Battista Della Porta in the frontispiece of the 1677 Italian edition of his *Magia naturalis* (Della Porta, 1677) as shown here in figure 3. But this is not the point. The point is that the doubt, mildly expressed by Favaro and possibly by Riccardi, concerns exclusively the orientation of the two figures. As a matter of fact, neither Favaro nor Riccardi (or anyone else) ever suggested or intimated that, even reading “19” instead of “61”, the portrait shows Fontana’s appearance at that age, but only that he possibly was nineteen in 1608. In other words, also in this interpretation, like in the previous one, it is understood that the effigy shows Fontana as he looked at the time of publication of the *Observationes*.[5] Instead, as we have seen, Molaro, in order to justify a so marked difference between the hair (and not only) of the two portraits, is forced to interpret the inscription as: “Francesco Fontana, from Naples, [here shown as he looked] in 1608, at the age of 19 years old, when he invented a new [kind of] astronomical telescope”.

Even apart from the strained (and, in my opinion, utterly incorrect) interpretation of the inscription, many arguments can be adduced against Molaro’s thesis. For instance, a problematic issue seems to be the apparent ages of the two men: as a matter of fact, the one portrayed in the engraving does not look like nineteen years old, but rather more like a pretty mature person, as, among other things, the bags under his eyes seem to suggest. On the other hand, as noted by Molaro himself, if we assume that Fontana was nineteen years old in 1608, we must presume that “the possible Fontana in the painting of Ribera should be few years older” (Molaro, 2017b: 286); to be exact, since Ribera’s *Allegory* is believed to be painted between 1615 and 1616 or even earlier,[6] seven to eight years older, at most, than the Fontana in the self-portrait. Therefore the man of the painting should be aged between 26 and 27 or less. Honestly, it is difficult to believe that the one depicted by Ribera is a young man, in his mid-twenties, unless one wants doubt Ribera’s talent as a portraitist, just like Molaro doubts Fontana’s skill in self-portraiture, but, as we have seen, he considers —as I guess everyone considers— Ribera “one of the most talented painters of his times”.

Molaro, moreover, seems not to realize that some of his statements contradict each other. For instance, in the original version of his paper, entitled *Francesco Fontana and his astronomical Telescope* (available at: https://arxiv.org/abs/1704.05661; accessed: 23 February 2022), referring to the refined telescope the man of the painting supports with both hands, Molaro claims that he is holding it “on the wrong side” (p. 19). Actually, this statement is absolutely erroneous: the telescope is definitely of Italian workmanship, and, therefore, the objective lens is contained in the main tube, that, in the painting, is properly depicted at the far end opposite to that of the observer. Anyway, although it is difficult to believe that such a simple device was so mysterious to him that he could not understand the right way to watch through it, let us assume that Molaro is right. Then, if the man of the canvas really was Fontana, is it plausible that he held the telescope the wrong way round? It could be argued that Ribera depicted the telescope later, after he had finished painting the sitter, but also in that case, is it plausible that Fontana never saw again the canvas and did not warn him a so gross mistake?

Additionally, to explain why Fontana, who always lived in Naples, would have posed for *The Sight*, Molaro (2017b: 285) is compelled to hypothesize that the canvas was painted, or at least finished, in Naples. Albeit, as we have said, this could be possible, in Molaro's opinion this hypothesis would be corroborated not by stylistic arguments, but by the seascape which the window on the right of the painting overlooks (contrary to Rome, Naples is on the coast). However, the landscape outside the window, toward which the telescope is ideally pointed, does not seem a seascape. In fact, the presence of trees so close to the water (figure 4) reminds rather of a river (Tiber?) or possibly a lake scenery (and the region around Rome is rich of lakes); and even if it were, it should be construed not as a faithful representation of the real place where the painting was depicted, but rather as a symbolic element, dare I say "mandatory" in an allegory of sight, the sense of long distances and vast spaces par excellence. It is no coincidence that *The Sight* is the only painting of Ribera's series of the *Five Senses* which has a window, without which the view would have been restricted to the narrow space delimited by the two walls that forms the background of the painting. An analogous function is carried out by the high opening, round arch of the *Sense of Sight*, depicted by Brueghel the Elder (1568–1625) in the same years in which Ribera painted *The Sight*, and by the wide double round arch window, on the right of the painting, in the *Allegory of Sight*, depicted in c. 1660 by Jan Brueghel the Younger (1601–1678)

Finally, as for the alleged difficulty in matching an object, at that time, so valuable to a street man (Molaro 2017b: 286), an argument that I find somewhat weak in itself, the real truth is that everything in the man of the painting —including his strong and virile hands and his attire, so in contrast to the refined elegance of the doublet in Fontana self-portrait— remind us of a man of the people. Ergo, also in the present case, Ribera does not seem to have made an exception to his custom of choosing everyday sitters, depicted in their raw realism. On the other hand, all art historians agree on this point. For example, Nicola Spinosa (2006: 19) describes the sitter as "a grim character, with thin hair, large ears, the face burned by the wind and the sun, tattered clothes [...] —maybe a farmer, a meat or offal dealer or a grain merchant?— called to represent the sense of sight and who now [...] he is pensively holding, in his rough and swollen hands, a telescope of the finest workmanship [my translation]"; and Gianni Papi (2007: 165) refers to to him as "a physiognomy that is anything but intellectual, who handles the instrument [i.e. the telescope] as if it were a work tool or even a weapon [my translation]".

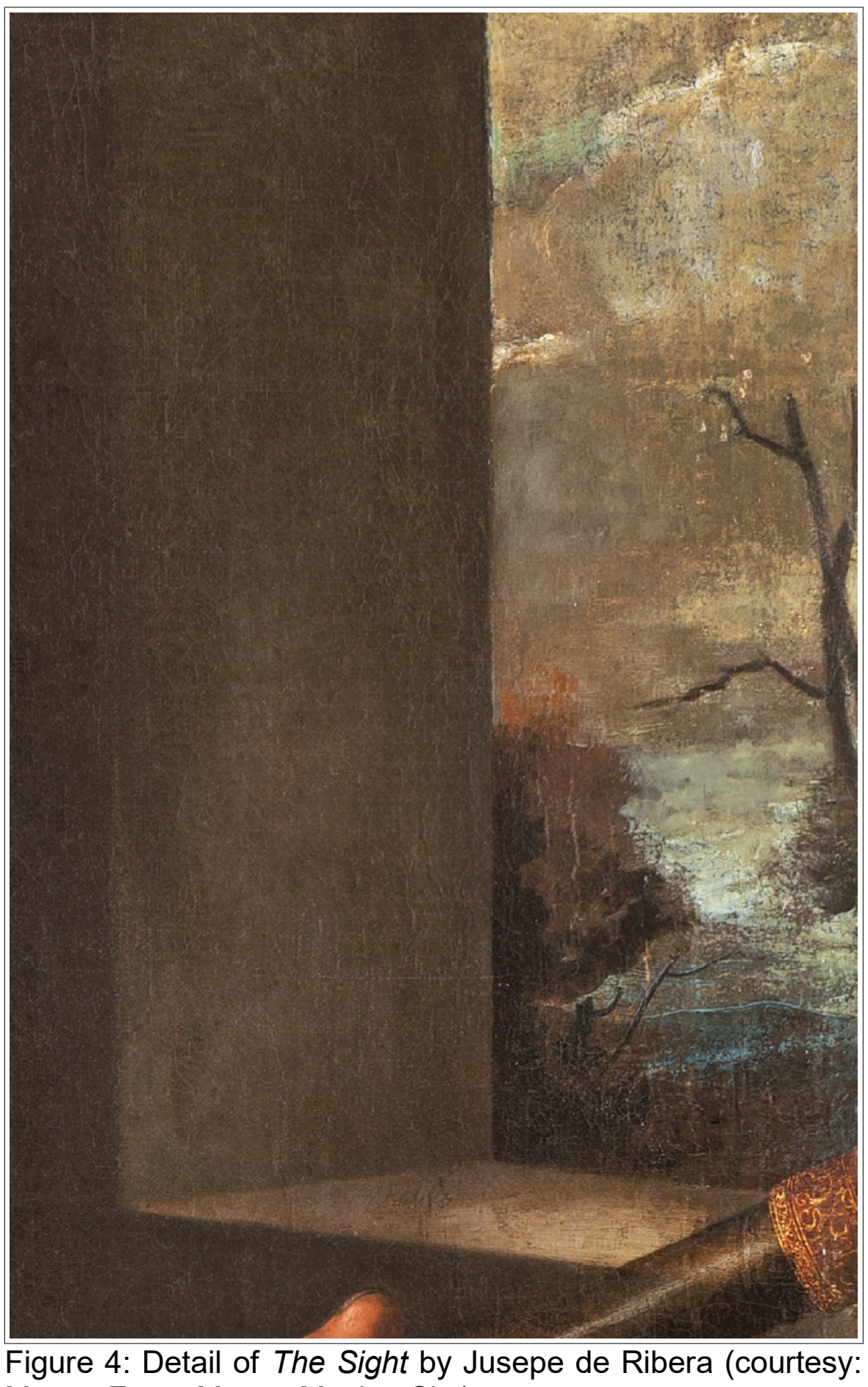

Figure 4: Detail of *The Sight* by Jusepe de Ribera (courtesy: Museo Franz Mayer, Mexico City)

## 2 THE BEGINNINGS OF FONTANA'S ACTIVITY

Up to this point, I have tried to demonstrate the total groundlessness of Molaro's claim substantially on the basis of arguments directly concerning the two portraits. I will now attempt to demolish his theory by means of a number of already known, so to speak, extra-pictorial sources, which Molaro seems to be unaware of. The alleged possibility that a great painter like Ribera deliberately wanted Francesco Fontana as a model for his *Allegory of Sight*, implies that the latter was already a famous maker of optical instruments —so famous as to be known even outside the restricted circle of the astronomers and naturalists— around 1615 or even earlier. As in a sort of *modus tollens*, proving the falsity of such an implication is equivalent to prove the falsity of the assumption, and, more importantly, this will allows us to come to a more trustworthy and less imaginary chronology of Fontana's activity.

First af all, we learn from Lorenzo Crasso (1666: 297) that Fontana tried to come into possession of Giovanni Battista Della Porta's optical tools, after the death of the latter in February 1615. But, if Fontana really already was a successful optician at that time, it seems rather improbable that he was trying to come into possession of Della Porta's utensils. Assuming that the episode is not spurious, it is much more reasonable to suppose that Fontana hoped to find out the actual or presumed optical secrets of the Neapolitan philosopher and/or that he wanted to get the tools to set up *ex novo* an optics workshop. Pamela Anastasio (1997: 652) seems to be of the same opinion when she claims that the attempt was evidently to obtain instruments and tools to start up his own business at that time. Moreover, she takes as an indirect chronological confirmation of this hypothesis a letter written by Evangelista Torricelli to Raffaello Magiotti on 6 February 1644, in which a telescope lens is described as "the best ever made by Fontana in a thousand glass over a period of 30 years" (*ibid.*).

Other indications —entirely consistent with the above-mentioned argument and much stronger— that around 1615, and still less earlier, Fontana was not so renowned as Molaro claims he was, are contained in three letters from Fabio Colonna —fellow citizen of Fontana and member of the Accademia dei Lincei— to Galileo. In the first one, dated 3 August 1613, Colonna writes: "I observed many sunspots and the spots of the Moon as well, even though *in Naples there is non one who can make perfect telescopes* [emphasis mine], therefore we could not see the new stars [i.e. the Galilean satellites of Jupiter]; three days ago I began to make by myself, to try, if I can, the convex clear without that little cloud; and I find many flaws both in the glasses and in their manufacture, and I commissioned some lenses eight and ten palms in diameter.[7] All of them turn out to be flawed, or dark" (Colonna, 1613a: 446). A few weeks later, in a letter dated 25 September of the same year, in which he reports on his progress on manufacturing telescope lenses, Colonna reaffirms that "in Naples there is no one who can be taught [to make telescopes], because there is no one who is both a theorist and a practician" (Colonna, 1613b: 464). Finally, two years later, on 14 August 1615, Colonna writes to Galilei that he is working on a new 14 palms telescope and regrets the lack of good glasses and good lathe turners, able to make copper molds to grind lenses (Colonna, 1615). In other words, in summer of 1615, Colonna complained about the utter lack of Neapolitan craftsmen able to produce lenses suitable to astronomical purposes.

Furthermore, there is a complete absence of any reference to Francesco Fontana in the correspondence of Italian astronomers and scholars of that decade. This is why Molaro is compelled to introduce the umpteenth *ad hoc* hypothesis that "it is quite possible [..] that Fontana's instruments reached the far courts in northern Europe even before other places in Italy" (Molaro, 2017: 284). But we wonder: did Fontana become known out of Italy even before than in his own Naples?! And, in any case, no mention by any European author is known in 1610s. In this regard, we can observe that in a letter, dated 25 marzo 1634 Christoph Scheiner, in response to Athanasius Kircher, who many years later, in 1646, would have placed Fontana together with Torricelli between the *praestantissimi artifices* (Kircher, 1671: 727), claims: "De Miraculo Neapolitano nihil scio" (Scheiner, 1634). So, still in 1634, in Vienna, one of the preeminent scholars of those years had not yet heard of Fontana's telescopes.

The lack of any reference to Fontana's activity persisted throughout the next ten years after Ribera painted *The Sight*, and, as a matter of fact, the first known allusion to Fontana is in a letter, dated 17 July 1626, from Colonna to Federeico Cesi (1585–1630), one of the founders of the Accademia dei Lincei:

> God willing, I will send Your Excellency a microscope, for which I am making the base and the screw tube, and that will be no more than four fingers in length, through which one can observe all day long without tiring the eyes, and it produces upright images: it has been invented by a friend, whom I am also helping to print his invention, since he wanted to make one [microscope] just like those of the Colognians [i.e. the brothers opticians Abraham and Jacob Kuffler], but, having failed to know its secret, kept investigating and discovered a better one (Colonna, 1626a).

Two months later, on 19 September, Colonna writes to Cesi:

> This friend [Francesco Fontana] has also invented another [kind] of *occhiale*, only one palm long, which produces upside-down images, but magnifies objects very much, and, what is most remarkable, it shows them so near that those which are as far away as a musket shot are seen close to the eyes (Colonna, 1626c).

The above-mentioned letters pose a number of interesting questions. First of all, why, in both of them, Colonna refers to Fontana as “un amico” [a friend] instead of calling him by name, as it might have been expected if the Neapolitan optician actually was already famous? Colonna, moreover, refers to Fontana as “the friend” at least in two more letters: one to Cesi, dated 22 August (Colonna, 1626b) of the same year, and one to Francsco Stelluti, dated 29 January of the next year (Colonna 1627). Furthermore, in the letter of 17 July 1626, Colonna claims that he is helping Fontana “to print his [alleged] invention”, which suggests that Fontana’s microscopes were still totally unknown in Naples.

The second and much more interesting question is: if Fontana were already renowned for his telescopes since mid-1610s, why, ten or more years later, in September 1626, did he show Colonna an instrument only one-palm in length? And —which is the specular question— why Colonna, who had been making telescope by himself for a dozen years and certainly was not unfamiliar with practical optics, was so impressed with a so small instrument? Why he makes no mention of some other larger instruments by Fontana? Probably because the small one-palm telescope that Fontana showed to Colonna was one of the first he had made, if not even the first-ever one, at least so high-performance.

From the above I think it can be concluded beyond a reasonable doubt that Fontana’s optical business began much later than the mid-1610s, and that it started about ten years later, around 1625 or shortly earlier, probably as a maker of microscopes. On the other hand, this dating of Fontana’s beginnings is much more consistent with the chronology of his successive accomplishments: the first 8-palm telescopes in autumn 1629 (Colonna, 1629), 14-palm in summer 1638[8] and a 22-palms the following year (for a chronology of Fontana’s production see Del Santo, 2009).

## 3 CONCLUSIONS

In the light of all the above-mentioned arguments it seems clear that the whole story narrated by Molaro is devoid of any reasonable foundation. In this story a Neapolitan craftsman (whom Molaro considers also a talented astronomer), named Francesco Fontana, would have invented, made and used the astronomical telescope before it was theorized, in 1611, by Kepler, and possibly even before Harriot and Galileo (“Allowing for some time to improve the quality of the lenses, [...] the year 1608 does not seem implausible as the birth-date of Fontana's [i.e. Keplerian] telescope, though it is based only on his own word” (Molaro, 2017b: 275)). Due to quality of his telescopes, Fontana would have achieved, in the city of Naples and in Europe before than in the rest of Italy, since mid-1610s, such a reputation as an optician to induce one of the most important and influential painter of his time to portray him in one of his painting. However, as acknowledged also by Molaro, this narration is based only on Fontana’s own claims (in which Molaro puts too much faith[9]), and on numerous *ad hoc* assumptions, the most outlandish of which is that he would have used the far-fetched, ridiculous expedient to draw a sort of toupee on his own head to look like forty years younger. In short, it is kind of like what happens in the famous novel *The Picture of Dorian Gray*, by Oscar Wilde, but with roles reversed, in which the real Francesco Fontana ages while his portrait remains young!

As we have seen, against this reconstruction there are several counter-arguments directly concerning the two portraits, and, at the same time, there is no concomitant evidence supporting it. On the contrary, there is a number of very persuasive arguments, based on epistolary sources (or on the absence thereof), which reveal that, around 1615, Fontana was still totally unknown, even in the inner circle of the specialists, both in Italy and abroad.

If to all this we add the alleged, but absolutely untenable, use of Keplerian telescopes by Fabio Colonna from as early as October 1614, maintained in some recent papers by Mauro Gargano (2019: 54; Gargano’s statement has been exhaustively discussed and demolished in Del Santo, 2021), and the presumed (and equally untenable) production of this kind of telescope since 1617 or even earlier (Molaro & Selvelli, 2011: 331–32), what emerges is a seemingly consistent, but actually totally fictional picture. In other words —also judging by the space that it has found in peer-reviewed journals and international conferences apparently, due to a sort of Gresham’s law concerning human mind, by virtue of which bad ideas drive out good—we are in danger of (re)writing a totally fabricated historiography not only of Fontana’s work, but of the whole early history of the telescope.

## 4 ACKNOWLEDGMENTS

I would like to thank Dr Giorgio Strano for reading the manuscript of the present paper and for providing helpful suggestions, and Dr Carlo Avilio for the valuable information furnished.

1 Aactually Molaro had already raised the possibility of this identification the previous year (see Molaro, 2017a: 231).

2 The book to which Molaro refers is Francesco Fontana, *Novæ cœlestium terrestriumque rerum observationes, Et fortasse hactenus non vulgatæ*, Neapoli, apud Gaffarum, 1646.

3 Actually, Fontana's portrait of the *Observationes* is a copper engraving to which Molaro always refers as a self-portrait, but, although unsigned, it could be (and likely it was) made from an anonymous professional engraver. At any rate, from now on, I too shall refer to it as the 'self-portrait' for the sake of brevity.

4 Actually, in a footnote, copying the inscription surrounding Fontana's portrait, Favaro writes: "Franciscus Fontana Neapol. novi optici tubi astronomici inventor A, Dom. M.DC.VIII. Aet. suae 61 (o 19?)".

5 Is evident that the two interpretations give different years of birth: if, at the time of the publication of the *Observationes*, Fontana was sixty-one years old, he would be born in around 1585, while, if he was nineteen years old in 1608, he would be born in around 1589. However, both birthdates are consistent with the very few information we have about Fontana's life.

6 On the basis of stylistic considerations, Giovanni Papi (2011: 52) believes that the canvas was painted towards the end of his Roman period, while Nicola Spinosa (2011: 90) in the early Neapolitan period. Ribera moved to Naples in the middle of 1616. Alfonso Pérez Sánchez (Pérez Sánchez & Spinosa, 1992: 60) dates the series of the Five Senses, whose order of creation is unknown, between 1611 and 1615.

7 Naturally, such sizes do not refer to the diameter of the objective lens, but to twice its radius of curvature.

8 Actually, Fontana had already made a 15-palm Telescope in March 1637, but this instrument turned out to be of quite poor optical quality, since it did "not define well Jupiter's disk, showing it all fluffy" (Magiotti, 1637).

9 Molaro's arguments are often really very tenuous, and the way he treats historical sources is quite questionable: for example, when he states that "[t]here are no apparent reasons to question Father Zupus' declaration to have used Fontana's [Keplerian] telescope in 1614, since he was still alive when the book was published" (*Ibid.*).